# Network-Centric Quantum Communications with Application to Critical Infrastructure Protection

Richard J. Hughes, Jane E. Nordholt, Kevin P. McCabe, Raymond T. Newell, Charles G. Peterson and Rolando D. Somma

Los Alamos National Laboratory, Los Alamos, NM 87545, USA

## ABSTRACT

Network-centric quantum communications (NQC) - a new, scalable instantiation of quantum cryptography providing key management with forward security for lightweight encryption, authentication and digital signatures in optical networks – is briefly described. Results from a multi-node experimental test-bed utilizing integrated photonics quantum communications components, known as QKarDs, include: quantum identification; verifiable quantum secret sharing; multi-party authenticated key establishment, including group keying; and single-fiber quantum-secured communications that can be applied as a security retrofit/upgrade to existing optical fiber installations. A demonstration that NQC meets the challenging simultaneous latency and security requirements of electric grid control communications, which cannot be met without compromises using conventional cryptography, is described.

# Network-Centric Quantum Communications with Application to Critical Infrastructure Protection

Richard J. Hughes, Jane E. Nordholt, Kevin P. McCabe, Raymond T. Newell,
Charles G. Peterson and Rolando D. Somma
Los Alamos National Laboratory, Los Alamos, NM 87545, USA

EXTENDED ABSTRACT

Trends in networking are presenting new security concerns that are challenging to meet with conventional cryptography, owing to constrained computational resources or the difficulty of providing suitable key management. In principle, quantum cryptography [1] with its forward security and lightweight computational footprint [2] could meet these challenges, provided it could evolve from the current point-to-point instantiations to a form compatible with multi-node networks [3, 4]. Trusted quantum key distribution (QKD) networks [5, 6, 7, 8] based on a mesh of point-to-point links lack scalability, require dedicated optical fiber, are expensive and not amenable to mass production, only provide one of the cryptographic functions (key distribution) needed for secure communications, and so have elicited limited practical interest. We have experimentally demonstrated a new, scalable approach to quantum information assurance called network-centric quantum communications (NQC), and have shown that it can solve new network security challenges in the critical infrastructure control sector, in particular.

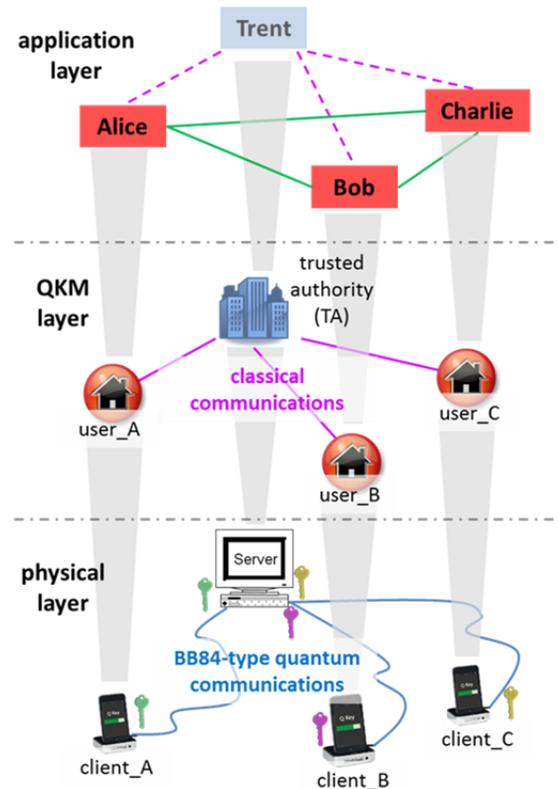

Fig. 1. Network-centric quantum communications (NQC) architecture. (See text for details.)

In NQC, BB84-type quantum communications (QC) [1, 9] between each of N client nodes and a central server node at the physical layer support a quantum key management (QKM) layer, which in turn enables secure communications functions (confidentiality, authentication and non-repudiation) at the application layer between ~ $N^2$ client pairs [10]. (See Fig. 1.) The NQC "hub-and-spoke" topology is widely encountered in optical fiber networks, and permits a hierarchical trust architecture that allows the server (the "hub") to act as the trusted authority (TA, "Trent") in cryptographic protocols for quantum authenticated key establishment (QAKE). (This avoids the poor scaling of previous approaches that require a pre-existing trust relationship between every pair of nodes.) By making Trent a single multiplexed QC receiver, and the client nodes (Alice, Bob, Charlie, etc.) QC transmitters, NQC amortizes the cost and complexity of one of the most demanding QC components – the single-photon detectors – across multiple network nodes [11]. In this way the NQC architecture is scalable in terms of both quantum-physical resources and trust.

The photonic phase-based qubits typically used in optical fiber QC require interferometric stability and inevitably necessitate bulky and expensive hardware. Instead, for NQC we use polarization qubits, allowing the QC transmitters – referred to as QKarDs - to be miniaturized and fabricated using integrated photonics methods [12]. This opens the door to a manufacturing process with its attendant economy of scale, and ultimately much lower-cost QC hardware. Our first-generation, modularly-integrated QKarD is a fiber-coupled device, similar in size to an electro-optic modulator, which incorporates a distributed feedback laser and modulator, that produces short ($< 1$ns), 1,550-nm weak-coherent (mean-photon number $< 1$), decoy-state, BB84 polarization-state light pulses at a 10 MHz repetition rate in our NQC test bed. (See Fig. 2.) This device is produced by a small-scale manufacturing process and could be incorporated into end devices. (Our next-generation, monolithically-integrated QKarD will be an order of magnitude smaller in each linear dimension.)

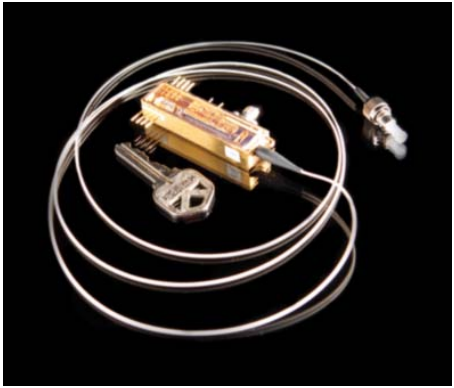

Fig. 2. Generation 1 quantum smartcard or QKarD. (See text for details.)

Trent, the NQC receiver, incorporates passive polarization analysis, with four cooled Geiger-mode InGaAs single-photon detectors, one for each BB84 state, operated at a typical detection efficiency of ~ 15%, a dark count probability of ~ $10^{-5}$ per 1-ns time-slot, and an after-pulse blocking time of ~ 50 μs applied to all four detectors after any detection. In our NQC test bed, which we have operated continuously for 2.5 years, we have time-multiplexed Trent with three QKarD transmitters, Alice, Bob and Charlie, over 50 km of single-mode fiber; larger numbers of clients could be accommodated with a combination of temporal and wavelength multiplexing. Fiber birefringence necessitates polarization tracking and compensation [13], which in our test bed is performed in-band (1,550-nm) at Trent, temporally multiplexed with the BB84 signals, resulting in a QC duty cycle of 20%. (When implemented out-of-band we expect our novel tracking scheme [14] will compensate even the much higher polarization rates that will be encountered in challenging environments such as aerial fibers [15].)

We operated several BB84-type QC protocols on each client-Trent link in our test bed. We have implemented three-level, decoy-state [16, 17, 18] verifiable quantum secret sharing (QSS) [19] over fiber spans as large as 50km between a client and a single, logical Trent node, which is comprised of two physical nodes: an intermediate, integrated-photonics BB84 modulation node and a BB84 receiver node [20]. In this way a client's key is split between the two physical parts of Trent, so that compromise of either one of the two would not compromise the key [21]. (Splitting of the key into additional shares would be possible with more intermediate nodes.) However, for brevity we will describe a simpler test bed configuration in which a three-level decoy-state QKD protocol, with mean-photon number signals of ~ 0, 0.1 and 0.5, is performed on the client-Trent links.

Each client uses true random numbers for decoy selection and BB84 basis and bit values. (In future versions we will use an integrated photonics quantum random number generator known as Velocirandor, which is capable of rates well in excess of the 5 Gbps already demonstrated [22].) Following sifting and shuffling, QKD reconciliation is accomplished with LDPC codes, privacy amplification [23] with Toeplitz hashing [24], and authentication [25] with cryptographic CRC hash functions. (We have also demonstrated that a QKD authentication key can be generated using a quantum identification protocol, QID [9], rather than pre-placing it. Further, QID provides a mechanism for client device enrollment and revocation [26, 27].) In one instantiation we have implemented the bi-directional classical communications parts of the QKD protocol using commodity 1,310-nm 1000Base-LX optical communications on the same fiber, with the majority of the bandwidth available for applications, forming a standalone quantum-secured communications (QSC) system. (Our QSC co-existence/filtering architecture permits sufficiently low sifted bit error rates (BER) for QKD up to 50 km of fiber.) The QKD data processing is handled by PCs at each node, with Trent's software architecture capable of accommodating up to 100 client QKarDs on that platform, and more than 1,000 with server-class hardware [28]. We have demonstrated that the processing software in each client can be transitioned to a field-programmable gate array (FPGA), providing further miniaturization for use cases such as hand-held device security.

Following the generation of shared secret keys between each client and Trent using QKD (or QSS), NQC protocols at the QKM layer provide keys for client-to-client secure communications at the application layer [27, 29]. For example, suppose Alice requires a secret key to encrypt a message to Bob using the AES algorithm [30], and/or authenticate the message using the HMAC algorithm [31]. (Other symmetric key encryption/authentication algorithms could be used instead.) For definiteness, we assume that this is a 256-bit key. First, Alice and Trent parse the results of their QKD session into three 256-bit keys: $K_A$, $L_A$ and $M_A$. Similarly, Bob and Trent parse the results of their QKD session into 256-bit keys $K_B$, $L_B$, and $M_B$. The key $K_B$ is going to become the key that Alice will use for securing her communications *to* Bob using the following QAKE procedure, reminiscent of the Leighton-Micali (LM) protocol [32]. Trent publishes the non-secret pair key, $P_{AB} = L_A \oplus K_B$ and confirmation key, $A_{AB} = H[K_B \parallel M_A]$, where H is a cryptographic hash function such as SHA-256 [33], and "$\parallel$" denotes concatenation of bit strings. Alice can then obtain the key $K_B$ that she needs to communicate to Bob by calculating $P_{AB} \oplus L_A$. She can confirm that she has the correct key by calculating $H[P_{AB} \oplus L_A \parallel M_A]$ and verifying that it equals the value $A_{AB}$ provided by Trent. This second, key confirmation step is essential for security against malicious manipulation of $P_{AB}$. Similarly, Bob would derive the distinct key, $K_A$, for securing his messages to Alice. Note that important security advantages of QAKE in contrast to LM are that: it does not require pre-placement of long-term secret keys shared by a client and Trent; and each new client-to-client key has no algorithmic heritage in any previous key. Trent could even go offline after publishing a look-up table of clients' pair keys and confirmation keys, and the clients would still be able to execute the above QAKE protocol using previously-generated, and securely-stored quantum keys.

A quantum digital signature (QDS) protocol [27] allows Alice to sign messages to Bob based on the Winternitz one-time signature (WOTS) scheme [34], which uses repeated cryptographic hash evaluations to calculate and verify signatures. Alice requires a secret signing key, S, to calculate her signature, and Bob requires Alice's non-secret signature verification key, V. In QDS Alice's S key is one of the keys she shares with Trent through QKD. Trent can calculate V from S, and provide it to Bob, authenticated with one of the Bob-Trent QKD keys. When Bob receives the signed message from Alice he can then verify her signature using V. Each Alice-Bob message requires a new signing key, which in QDS is generated from Alice-Trent QKD. This is an important advantage of QDS relative to conventional WOTS schemes that require pre-computation and pre-distribution of large Merkle trees for verifying signatures on multiple messages. As with QAKE, QDS has forward security, providing fresh signing keys instead of the pre-placed, algorithmically-related keys of the analogous classical protocols. With the inclusion of the digital signature capability, NQC can provide the full functionality of a public key infrastructure [27].

Energy delivery systems such as the SmartGrid have requirements for the secure, low-latency (~ few ms) delivery of data from, and commands to, control devices [35]. In the electricity transmission sector there is for example, a low-latency data origin authentication [36] requirement for the multicast of phasor measurement unit (PMU) data to phasor data concentrators (PDC) [37]. Analyses have shown that these two requirements cannot be met simultaneously with present-day cryptography on commodity processors [38], without compromising either the security assurances or the quality-of-service: RSA signatures [39] are too slow, for example; while message authentication codes using a common group key are vulnerable to compromise of a single node. (Similar security/latency requirement mismatches exist throughout the electric grid.) Newer, lightweight cryptography could meet the requirement, but with conventional approaches the key management would be impractical [38]. However, optical fiber is widely-deployed throughout the electric grid, and so NQC with its essentially unconstrained ability to supply fresh, forward-secure keys on-demand is capable of supporting lightweight cryptography [40] for these applications. Furthermore, our single-fiber QSC configuration has the highly desirable feature that it can be inserted as a "bump-in-the-wire" security retrofit into existing grid communications. In December 2012 we demonstrated these advantages of NQC in the US Department of Energy's Trustworthy Cyber Infrastructure for the Power Grid (TCIPG) test bed at the University of Illinois Urbana-Champaign (UIUC) [41]. (See Fig. 3.) With the TCIPG test-bed configured to simulate both normal and fault conditions in the electric transmission grid, our QSC system secured the PMU-PDC communications over a 25-km fiber link with only ~ 125 μs of additional latency, exceeding requirements by almost two orders of magnitude [42]. This result convincingly demonstrates the value that NQC offers for solving the pressing cyber security challenges of critical infrastructure protection. Other NQC use cases include: handheld device security; enterprise network security; and secure cloud computing.

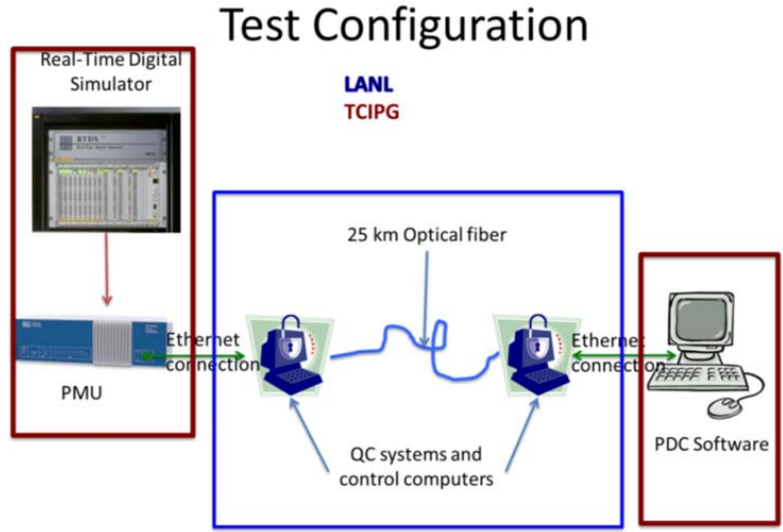

Fig. 3. Test configuration using the Los Alamos (LANL) QSC system to secure PMU control commands and data. The TCIPG test bed provided a real-time digital (power) simulator, PMU, and PDC software to control and display data from the PMU. (See text for details.)

**Acknowledgements**

The Los Alamos Laboratory Directed Research and Development (LDRD) program supported a portion of the research described in this paper. The support of the Department of Energy's Cybersecurity for Energy Delivery Systems (CEDS) program, and Dr Carol Hawk and Dr Diane Hooie in particular, is gratefully acknowledged. We thank William Sanders, Jeremy Jones and Tim Yardley of UIUC for their hospitality in facilitating the QSC demonstration at their TCIPG test bed. It is a pleasure to thank Joysree Aubrey, Brian Bluhm, Geoff Burdge, Tom Chapuran, Harald Dogliani, Bob Ecke, G. Andrew Erickson, Andrew W. Erickson, Carl Hauser, Ryan Kennedy, Marcus Lucero, Mike Mosca, Alex Rosiewicz, Masahide Sasaki, Rhett Smith, James Thrasher, Stephanie Wehner and Jon Yard for helpful discussions.

## References

1. C.H. Bennett and G. Brassard, "Quantum Cryptography: Public Key Distribution and Coin Tossing", Proc. of the IEEE Int. Conf. on Computers, Systems and Signal Processing, 175 (1984).
2. R. J. Hughes and J. E. Nordholt, "Refining Quantum Cryptography", Science **333**, 1584 (2011).
3. T. E. Chapuran et al., New J. Phys. **11**, 105001 (2009).
4. N. A. Peters et al., New J. Phys. **11**, 045012 (2009).
5. C. Elliott, New J. Phys. **4**, 46.1 (2002).
6. M. Peev et al., New J. Phys. **11**, 075001 (2009).
7. F. Xu et al., Chinese Science Bulletin **54**, 2991 (2009).
8. W. Maeda et al., IEEE J. Sel. Top. QE **15**, 1591 (2009).
9. I. Damgaard et al., Lect. Notes. Comp. Sci. **4622**, 342 (2007).
10. R. J. Hughes et al., "Secure multi-party communication with quantum key distribution managed by trusted authority", World Intellectual Property Organization (PCT) application WO 2012/044855, published April 5, 2012.
11. J. E. Nordholt et al., "Quantum key distribution using card, base station and trusted authority", World Intellectual Property Organization (PCT) application WO 2012/044852, published April 5, 2012.
12. J. E. Nordholt et al., "Quantum communications system with integrated photonics devices", US Patent application 61/684,502 (August 30, 2012).
13. J. E. Nordholt et al., "Polarization tracking system for free-space optical communication, including quantum communication", US Patent Application US20130083925, World Intellectual Property Organization (PCT) application WO 2013/048671, published April 4, 2013.
14. J. E. Nordholt et al., "Great circle solution to polarization-based quantum communication (QC) in optical fiber", US Patent Application US 20130084079, World Intellectual Property Organization (PCT) application WO 2013/048672, published April 4, 2013.
15. D. S. Waddy et al., Opt. Fiber Tech. **11**, 1 (2005).
16. W. Y. Hwang, Phys. Rev. Lett. **91**, 057901.1 (2004).
17. X. B. Wang, Phys. Rev. Lett. **94**, 230503.1 (2005).
18. J. W. Harrington et al., quant-ph/0503002 (2005).
19. M. Hillery et al., Phys. Rev. A **59**, 1829 (1999).
20. C. Schmidt et al., Phys. Rev. Lett. **95**, 230505 (2005); Phys. Rev. Lett. **98**, 028902 (2007).
21. P. D'Arco and D. R. Stinson, Lect. Notes. Comp. Sci. **2501**, 346 (2002).
22. J. E. Nordholt et al., "Ultra high speed quantum random number generator", US Patent Application 13/600,905 (August 30, 2012).
23. C. H. Bennett et al., IEEE Trans. Inf. Th. **41**, 1915 (1995).
24. H. Krawczyk, "LFSR-based Hashing and Authentication", Lect. Notes Comp. Sci. **839**, 129 (1994).
25. M. N. Wegman and J. L. Carter, "New Hash Functions and Their Use in Authentication and Set Equality", J. Comp. Sys. Sci. **22**, 265 (1981).
26. R. J. Hughes et al., "Multi-factor authentication using quantum communication", US Patent application 61/695,190 (August 30, 2012).
27. R. J. Hughes et al., "Quantum Key Management", US Patent Application US 20130083926, World Intellectual Property Organization (PCT) application WO 2013/048674, published April 4, 2013.
28. J. E. Nordholt et al., "Scalable software architecture for quantum cryptographic key management", US Patent application 61/693,131 (August 30, 2012).

29. C. Blundo and P. D'Arco, Lect. Notes. Comp. Sci. **2946**, 44 (2004).
30. "NIST-FIPS-197: Announcing the Advanced Encryption Standard (AES)" National Institute of Standards and Technology (NIST), 2001.
31. "NIST-FIPS 198: The Keyed-Hash Message Authentication Code (HMAC)", National Institute of Standards and Technology (NIST), 2002.
32. T. Leighton and S. Micali, Lect. Notes. Comp. Sci. **773**, 456 (1994).
33. "NIST-FIPS-180-4: Secure Hash Standard (SHS)" National Institute of Standards and Technology (NIST), 2001.
34. R. C. Merkle, "A Certified Digital Signature", Lect. Notes Comp. Sci. **435**, 218 (1990).
35. National Institute of Standards and Technology "Guidelines for Smart Grid Cyber Security: Vol. 1", NISTIR 7628, U.S. Department of Commerce, 2010.
36. For a survey, see Y. Challal et al., "A Taxonomy Of Multicast Data Origin Authentication: Issues And Solutions", IEEE Communications Surveys and Tutorials **6**(3), 34 (2004).
37. "IEEE Standard for Synchrophasors for Power Systems", IEEE C37.118 (2005).
38. C. H. Hauser et al., "Evaluating Multicast Message Authentication Protocols for Use in Wide Area Power Grid Data Delivery Services", Proceedings IEEE 45th Hawaii International Conference on System Sciences (HICSS), 2151 (2012).
39. R. L. Rivest et al., "A Method for Obtaining Digital Signatures and Public-Key Cryptosystems", Communications of the ACM **21**(2) 120 (1978).
40. For a survey see, for example T. Eisenbarth et al., IEEE Design and Test **24**, 522 (2007).
41. http://tcipg.org/research_Testbed
42. A video of the demonstration is available at: https://uofi.box.com/qkd-demo